\documentclass[conference]{IEEEtran}
\usepackage{siunitx}
\usepackage{cite}
\usepackage{textcomp}
\def\BibTeX{{\rm B\kern-.05em{\sc i\kern-.025em b}\kern-.08em
    T\kern-.1667em\lower.7ex\hbox{E}\kern-.125emX}}

\usepackage{goldschmidt_ieee}
                    
\graphicspath{{figs/}} 

\begin{document}

\title{Using optimal control to guide neural-network interpolation of continuously-parameterized gates}

\author{
    \IEEEauthorblockN{
        Bikrant Bhattacharyya\IEEEauthorrefmark{2},
        Fredy An\IEEEauthorrefmark{2},
        Dominik Kozbiel\IEEEauthorrefmark{2},
        Andy J. Goldschmidt\IEEEauthorrefmark{1}\IEEEauthorrefmark{3},
        Frederic T. Chong\IEEEauthorrefmark{3}
    }
    \IEEEauthorblockA{\IEEEauthorrefmark{2}Illinois Mathematics and Science Academy, Aurora, IL 60506}
    \IEEEauthorblockA{\IEEEauthorrefmark{3}Department of Computer Science, University of Chicago, Chicago, IL 60637}
    \IEEEauthorblockA{\IEEEauthorrefmark{1}Corresponding author: andygold@uchicago.edu}
}

\maketitle
\global\csname @topnum\endcsname 0
\global\csname @botnum\endcsname 0

\thispagestyle{plain}
\pagestyle{plain}

\begin{abstract}
Control synthesis for continuously-parameterized families of quantum gates can enable critical advantages for mid-sized quantum computing applications in advance of fault-tolerance.
We combine quantum optimal control with physics-informed machine learning to efficiently synthesize control surfaces that interpolate among continuously-parameterized gate families.
Using optimal control as an active learning strategy to guide pretraining, we bootstrap a physics-informed neural network to achieve rapid convergence to nonlinear control surfaces sufficient for our desired gates. 
We find our approach is critical for enabling an expressiveness beyond linear interpolation, which is important in cases of hard quantum control.
We show in simulation that by adapting our pretraining to use a few reference pulse calibrations, we can apply transfer learning to quickly calibrate our learned control surfaces when devices fluctuate over time.

We demonstrate synthesis for one and two qubit gates with one or two parameters, focusing on gate families for variational quantum algorithm (VQA) ansatz. By avoiding the inefficient decomposition of VQA ansatz into basis gate sets, continuous gate families are a potential method to improve the noise robustness of VQAs in the near term.
Our framework shows how accessible optimal control tools can be combined with simple machine learning to enable practitioners to achieve 3x speedups for their algorithms by going beyond the standard gate sets.
\end{abstract}


 \begin{figure}[t]
    \centering
    \includegraphics[width=\columnwidth]{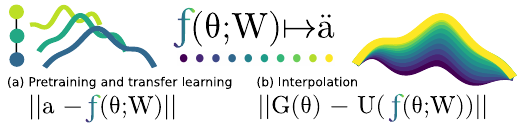}
    \caption{
        A physics-informed neural network (NN), $f(\cdot\,; \mathbf{W})$, is used to map from gate parameters, $\boldsymbol{\theta}$, to control accelerations, $\ddot{\mathbf{a}}$. (a) The NN actively learns during a pretraining phase, guided by pulse designs from a coordinated quantum optimal control problem.  (b) The network is further trained on a fidelity-based loss in order to learn to interpolate. The result is an efficient and compact representation of high-fidelity controls that realize a continuously-parameterized quantum gate family. After training, transfer learning can be applied in (a) using experimentally-calibrated pulses to act as new guides, thereby recalibrating the entire learned gate family.
    }
    \label{fig:rought_main}
\end{figure}

\section{Introduction} \label{sec:introduction}
During a quantum algorithm, the state of the quantum computer evolves continuously in time. If the system is isolated from its environment, the dynamics are described by the Schr\"{o}dinger equation, $i |\dot{\boldsymbol{\psi}}\rangle = \mathbf{H} \ket{\boldsymbol{\psi}}$: Hamiltonians generate continuous rotations, $\mathbf{U}(t) = \exp(-i \mathbf{H} t)$, of the quantum register, $\ket{\boldsymbol{\psi}}$. In practice, the analog rotations describing the algorithm are compiled into a finite, universal gate set of one and two-qubit operations~\cite{cross2017open,smith2016practical,haner2018software}. The purpose of a gate set is to simplify the number of accurate operations the computer needs to perform, and to digitize any analog errors to enable their correction. Gate sets are necessary for fault-tolerant quantum computation. However, for promising near-term applications like quantum chemistry, restriction to a gate set means multi-qubit or parametric interactions must be rewritten in terms of standard gate sets---a potentially inefficient transcription~\cite{preskill2018quantum}.

In light of the finite decoherence times of mid-sized quantum computers--the ones that are too small or noisy to realize quantum error correction--moving to direct analog synthesis of multi-qubit operations can provide critical improvements for important near-term quantum algorithms~\cite{shi2019optimized}. In particular, Variational Quantum Algorithms (VQAs) use the output of parameterized quantum circuits to optimize an objective within a hybrid quantum-classical optimization loop~\cite{cerezo2021variational}. Indeed, VQAs can benefit from direct synthesis; however, parameters are updated dynamically, and independently synthesizing each operation can become prohibitive.  Analog gate families, which restrict the allowed operations to certain continuously parameterized gates, offer a practical middle ground between direct synthesis and digital gates~\cite{abrams2020implementation,foxen2020demonstrating,lacroix2020improving,shi2021simulating}. To select one example, a demonstration of high-fidelity continuous gate families for superconducting qubits enabled a 3x improvement in circuit depth over a standard gate set~\cite{foxen2020demonstrating}. Beyond improving depth, gate families can be chosen to respect known problem symmetries or to reduce circuit over-parameterization in VQAs, resulting in lower resource overheads and improved algorithm performance~\cite{grimsley2019adaptive,gard2020efficient,shi2021simulating,bertels2022symmetry}. In particular, we focus on the operators from the \textsc{QEB-ADAPT-VQE} operator pool \cite{yordanov2020efficient}. The \textsc{QEB-ADAPT-VQE} algorithm constructs a problem specific ansatz to find the ground state of a molecular Hamiltonian. We synthesize pulses for a two qubit, single parameter \textsc{QEB-ADAPT-VQE} gate, resulting in circuits that are $2.78-3.28\times$ shorter than the standard basis gate decompositions.

Frameworks have been proposed for synthesizing continuously-parameterized families of gates~\cite{sauvage2022optimal, preti2022continuous, chadwick2023efficient}. All approaches rely on continuous interpolation of system Hamiltonians between reference parameters, so that a high-fidelity gate can be achieved without needing synthesis at each parameter value. In general, achieving any gate on a quantum computer requires precise control of the Hamiltonians generating the device dynamics. Physically, controls ($\mathbf{a}_t$) are electromagnetic fields that couple to the quantum system's Hamiltonians; mathematically, this is a \textit{bilinear control system} in states and controls~\cite{goldschmidt2021bilinear}: $ i |\dot{\boldsymbol{\psi}}\rangle = \mathbf{H}(\mathbf{a}_t) \ket{\boldsymbol{\psi}} = (\mathbf{H}_0 + \sum_{j} \text{a}_{jt} \mathbf{H}_j ) \ket{\boldsymbol{\psi}}$. The time-ordered integral of the Schr\"{o}dinger equation returns the resulting gate. In practice, simple gates may be designed analytically; otherwise, numerical optimal control algorithms are written to solve for control fields sufficient to achieve desired quantum operations~\cite{koch2022quantum}. In particular, we are interested in achieving efficient, minimum time gates under practical hardware constraints for our VQE application; the impact of these constraints result in hard quantum control problems, with the potential for suboptimal local minima in the nonlinear cost landscape over the controls~\cite{koch2022quantum}. 

Broadly speaking, past frameworks for synthesizing continuously-parameterized families of gates fall into two camps. The first idea is to use an artificial neural network (NN) to learn a mapping from parameters to controls; continuity of the network enables interpolation, which can be arbitrarily expressive~\cite{preti2022continuous, sauvage2022optimal}. The second approach is to iteratively solve optimal control problems at reference parameters while regularizing nearby controls to remain close during re-optimization~\cite{chadwick2023efficient}. In the latter, intermediate values are obtained with simple linear interpolation, saving the computational overhead of training a network.

We propose a unified framework for synthesis of continuous gate families that combines the merits of the previous approaches and overcomes their limitations. To efficiently train our NN, we use optimal control to design an \textit{active learning} scheme~\cite{cohn1996active}, which allows us to significantly boost training efficiency (achieving up to a $7\times$ improvement, see Table \ref{tab:table1}). We refer to this phase as our \textit{pretraining}. In our framework, we choose to implement our NN in a physics-informed way, thereby implicitly maintaining the structure of our desired optimal control solutions throughout training. By retaining a NN representation of the quantum gate family, we are also able to perform \textit{transfer learning} of the network weights for the purpose of calibration; our approach integrates with any existing calibration routine, and only requires the experimentalist to obtain a few reference pulses in order to recalibrate entire gate families. To the best of our knowledge, this is the first example of applying transfer learning directly on calibrated controls. In our pretraining phase, we also demonstrate the advantages offered by a simple and flexible approach to quantum optimal control called Pad\'{e} integrator direct collocation (Piccolo)~\cite{trowbridge2023direct}; we introduce an accessible problem template---which we have integrated into the open-source software package \texttt{Piccolo.jl}~\cite{website2024piccolo}---that allows for efficient solutions to direct-sum quantum optimal control problems essential for our method, the methods of~\cite{chadwick2023efficient}, and other coordinated optimization problems in quantum control. 

In what follows, we first introduce our framework (Section~\ref{sec:methods}); we describe our physics-informed NN architecture and how we use it for interpolation through our novel pretraining and training phases in Sections~\ref{sec:interpolation}-\ref{sec:training}. With Section~\ref{sec:calibrating}, we emphasize the practicality of our framework by showing how we adapt our pretraining scheme to realize transfer learning for efficient calibration of entire gate families. In Section~\ref{sec:synth}, we demonstrate our approach on a collection of one and two parameter gate families involving one and two qubits. We select gates that are relevant for near-term quantum computing applications, and we conclude with calculations of the advantage that quantum gate families are able to provide for our representative \textsc{QEB-ADAPT-VQE} applications in Section~\ref{sec:advantages}.

\section{Methods} \label{sec:methods}
Our framework for realizing continuously-parameterized quantum gate families relies on the use of a neural network (NN) to store an efficient parameterized representation of the control solution. We optimize the network by following the procedure outlined here: 
\begin{itemize}[leftmargin=1cm]
    \item[\ref{sec:interpolation}] We introduce our NN, which predicts the accelerations using the gate parameters as the network inputs. Although the NN weights are randomly initialized, because our architecture predicts control accelerations, we already capture part of the problem structure.
    \item[\ref{sec:qc}] Pretraining control accelerations are generated using the \texttt{Piccolo.jl} framework for reference parameter values. We use regularization to coordinate the solutions via a graph-based quantum optimal control problem that we refer to as a \textit{direct sum problem}.
    \item[\ref{sec:pretraining}] The NN is \textit{pretrained} using the control accelerations from quantum optimal control at a set of reference gate parameters. We pretrain by minimizing the mean squared error between these pulse values and the network predictions.
    \item[\ref{sec:training}] The network is \textit{trained} to generalize to the entire surface. We train by minimizing a loss function consisting of a term corresponding to the infidelity measured on the dataset and a regularization term. Every epoch, a new dataset of parameter values is selected to update the network. 
\end{itemize}
At training convergence, the network compactly stores an interpolation of a continuously-parameterized family of quantum gates.

 \begin{figure}[t]
    \centering
    \includegraphics[width=\columnwidth]{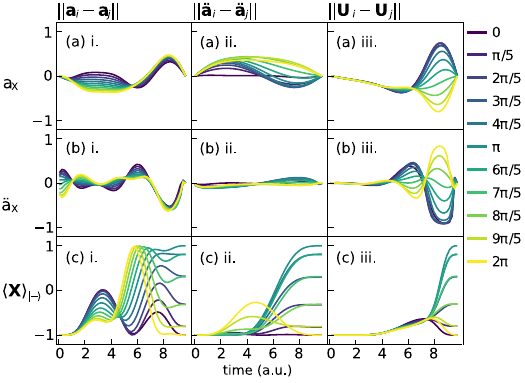}
    \caption{
    A comparison of regularization choices for~\eqref{eqn:direct-sum-problem} are shown for the $\mathbf{R_Z}(\theta)=\exp(-i \theta \mathbf{Z}/2)$ gate, via $\mathbf{H}(\boldsymbol{a}_t)=\text{a}_{\text{X}t}\mathbf{X}+\text{a}_{\text{Y}t}\mathbf{Y}$. The $\mathbf{X}$ control is shown in row one, the $\mathbf{X}$ acceleration in row two, and---as a proxy for the unitary---the $\mathbf{X}$ operator itself in row three (in the Heisenberg picture, started from its $-1$ eigenstate, $|-\rangle$). Colors are the solutions for reference angles $\theta \in [0, 2\pi]$. The controls for each reference angle are asked to satisfy a fidelity constraint of $0.9999$, while optimizing to minimize the distance between different optimization variables: the column titles indicate the chosen regularization, with subscripts $i$ and $j$ representing a nearest neighbor coupling between subproblems. From left to right, $(\text{a,b,c}) \text{i.}$ show the effect of minimizing the control distance over the entire trajectory, $(\text{a,b,c}) \text{ii.}$ the control acceleration distance, and $(\text{a,b,c}) \text{iii.}$ the unitary distance.
    }
    \label{fig:fig2_compare}
\end{figure}
\subsection{Interpolation using physics-informed machine learning} \label{sec:interpolation}
The goal of interpolation is to smoothly connect parameterized gates, $\mathbf{G}(\boldsymbol{\theta})$, by considering how the parameter relates to the controls, $\mathbf{a}(\boldsymbol{\theta})$. NNs can learn universal function approximations $\hat{\mathbf{f}}(\boldsymbol{\theta}; \mathbf{W}) \approx \mathbf{f}(\boldsymbol{\theta})$, by minimizing the network weights, $\mathbf{W}$, in order to match function outputs, $\mathbf{f}_j$; the objective is $\min_{\mathbf{W}} \sum_j ||\hat{\mathbf{f}}(\boldsymbol{\theta}_j; \mathbf{W}) - \mathbf{f}_j||$. NNs are highly effective at matching their predictions within the domain of a training set; most trouble arises when networks are asked to extrapolate. In our familial quantum control problems, we predict controls or their accelerations based on gate parameters, $\boldsymbol{\theta}$. Our problems have two structural advantages. First, we know that $\boldsymbol{\theta}$ reside in a direct product of intervals (e.g. $[0, 2\pi]^{\text{length}(\boldsymbol{\theta})}$), so we do not have to worry about extrapolation outside the training domain. Second, we are able to actively generate training data---we are not restricted to any samples $\boldsymbol{\theta}_j$, a priori. For these reasons, we find our problem to be ideally suited to interpolation by NN.

In our work, we use a feedforward NN to map gate parameters to \textit{piecewise-constant accelerations}: $\boldsymbol{\theta} \mapsto \ddot{\mathbf{a}}(\boldsymbol{\theta})$. We use acceleration in our interpolation for three reasons:
\begin{enumerate}
    \item[$\circ$]  Smooth controls can be guaranteed by returning bounded accelerations. This limits the risks due to \textit{overfitting oscillations}, akin to the oscillations seen in the Runge phenomenon\footnote{An over-parameterized (or under-regularized) network can overfit its training data, leading to high-variance predictions on unseen data points~(Ch.~5,~\cite{goodfellow2016deep}).}.
    \item[$\circ$] Two boundary conditions are necessary when applying Euler integration to obtain $\mathbf{a}$ from $\ddot{\mathbf{a}}$, so we can directly enforce $\mathbf{a}(0) = \mathbf{a}(T) = 0$.
    \item[$\circ$] The piecewise constant accelerations allow for greater control freedom; in the alternative where $t, \boldsymbol{\theta} \mapsto \mathbf{a}_t(\boldsymbol{\theta})$, the infinitely differentiable NN must remain smooth at every order of the time derivative.
\end{enumerate}
In contrast to our approach,~\cite{sauvage2022optimal} used a NN where $t, \boldsymbol{\theta} \mapsto \mathbf{a}_t(\boldsymbol{\theta})$, so that each timestep was treated as an input. This formulation relies on control outputs that vary smoothly over time and gate parameters; if these assumptions hold, a NN can readily attain an approximation with a few hidden layers. However, this precludes piecewise-constant controls or accelerations; in those cases, interpolation by NNs requires significant overhead in parameters due to the non-differentiable jumps. Indeed, even with a large parameter increase, the NN output might still be ill-behaved~\cite{selmic2002neural}. Hence, piecewise-constant functions should be treated as having independent outputs over all timesteps, e.g. $\boldsymbol{\theta} \mapsto \mathbf{a}(\boldsymbol{\theta}) = \begin{bmatrix} \mathbf{a}_1(\boldsymbol{\theta}) & \mathbf{a}_2(\boldsymbol{\theta}) & \cdots & \mathbf{a}_T(\boldsymbol{\theta}) \end{bmatrix}$. We find this modification to be essential for our network to predict piecewise-constant accelerations.

Interpolation by NN requires training on fidelity (discussed further in~\ref{sec:training}). This requires the computation of gradients through model rollouts, which can be a costly procedure~\cite{abdelhafez2019gradient}. The training cost is made worse if a randomly-initialized network must learn to interpolate while also learning to predict good controls. Alternatively,~\cite{chadwick2023efficient} demonstrated how a great deal of success could be had through structured and efficient linear interpolation schemes among controls that are close by in parameter space. The linear interpolation is much more efficient to compute than NN training, but it is a limiting approximation for three reasons. First, linear interpolation may not align with the true nonlinear control manifold. This is particularly relevant for hard quantum control problems like minimum time, which limit the available trajectories for realizing gates and otherwise disrupt the control landscape~\cite{koch2022quantum} (for an example, see Section~\ref{sec:two-parameter}; for additional discussion, see \ref{sec:qc:theory}). Second, optimizing directly for linear interpolation results in control shapes that work well for this purpose; however, restriction to the class of linearly-interpolatable controls may cause tension when other shape requirements, such as noise robustness, are considered~\cite{green2013arbitrary}. Third, we find the NN to be a useful compact representation of the correlations between gate parameters, allowing us to explore ideas like transfer learning (Section~\ref{sec:calibrating}). For these reasons, we seek a way to retain the arbitrary function approximation property of a NN when proposing a general framework for interpolation of gate families. In Section~\ref{sec:pretraining}, we build upon the lessons of~\cite{chadwick2023efficient} and use regularization to connect our control problems, introducing a quantum optimal scheme enabling generalizations of their approach along the way. Then, we use the regularized control solutions to guide our network via pretraining, allowing us to efficiently realize NN-based interpolation (Section~\ref{sec:training}).

\subsection{Quantum optimal control} \label{sec:qc}
Pad\'{e} integrator direct collocation (\textit{Piccolo}) is a particular type of optimal control algorithm known as a \textit{direct method}~\cite{trowbridge2023direct} (to be contrasted with an \textit{indirect method}, c.f. Section~\ref{sec:interpolation}). In a direct method, both the states and controls of the problem trajectory are optimization variables. The dynamics of the problem enter as constraints between timesteps, and are only enforced upon convergence. Indirect methods encompass most other quantum optimal control algorithms~\cite{khaneja2005optimal, muller2022one, machnes2018tunable, goerz2019krotov,petersson2021optimal,ball2021software}; in indirect methods, only the controls are optimization variables, and the state variables are accessed by rolling out (thereby enforcing) the dynamics. Direct methods have superior convergence properties because they can traverse the optimization landscape more efficiently by breaking constraints during optimization~\cite{trowbridge2023direct}.

For our purpose, we leverage the direct nature of \textit{Piccolo} for the added problem flexibility it provides. Denote $\mathbf{X} = \begin{bmatrix}  \boldsymbol{\chi}_1 & \boldsymbol{\chi}_2 & \dots & \boldsymbol{\chi}_T \end{bmatrix}$ as the state matrix and $\mathbf{A} = \begin{bmatrix} \boldsymbol{\alpha}_1 &  \boldsymbol{\alpha}_2 & \dots &  \boldsymbol{\alpha}_T \end{bmatrix}$ as the control (or actuation) matrix. All optimization variables for our direct optimization problem are collected into $\mathbf{Z} := \begin{bmatrix}  \boldsymbol{\zeta}_1 & \boldsymbol{\zeta}_2 & \dots & \boldsymbol{\zeta}_T \end{bmatrix} = \begin{bmatrix}\mathbf{X}; & \mathbf{A} \end{bmatrix} $. In our optimal control algorithm, we define the state and control more abstractly as augmented states and controls (benefits of this approach are discussed in~\cite{propson2022robust}); our state includes not only the (vectorized) unitary matrix, but also the control values and control velocities, so $\boldsymbol{\chi}_t = \begin{bmatrix} \vec{\mathbf{U}}_t; & \mathbf{a}_t; & \dot{\mathbf{a}}_t \end{bmatrix}$. The control matrix contains the accelerations of the controls and perhaps even the timesteps of the problem, $\boldsymbol{\alpha}_t = \begin{bmatrix} \ddot{\mathbf{a}}_t; & \Delta t_t \end{bmatrix}$. By using piecewise-constant acceleration as our control input, we are able to guarantee continuously differentiable pulse shapes without introducing further restrictions.

A key feature of our approach, which we inherit from~\cite{chadwick2023efficient}, is the use of coordinated optimizations among sampled parameters and their corresponding gates. We trace the idea of coordinated optimization to the classic idea of sampling-based learning control (applied to robust quantum optimal control in~\cite{dong2015sampling}), which solves a control problem over a representative ensemble of parameterized models. Interpolation over pulses from parameterized Hamiltonians for the purpose of robust control was also discussed in~\cite{luchi2023control}. A study of coordinated optimization among time-parameterized Hamiltonians for adiabatic quantum control can be found in~\cite{brif2014exploring}.

\subsubsection{Unitary smooth pulse problem template}
When optimizing for parameterized gate families, we start from a set of representative parameters and the corresponding gates, $\mathbf{G}(\boldsymbol{\theta}_j)$ for $j=1,2,\dots,S$. We solve independent, model-based optimization problems to achieve high-fidelity gates for each parameter. These subroutines are a core feature of Piccolo, and they follow the unitary smooth pulse problem template,
\begin{align} \label{eqn:smooth-pulse-problem}
    \arg \min_{\mathbf{Z}_{j}} & |1 - \mathcal{F}(\mathbf{Z}_{j}, \mathbf{G}_j)| + \frac{1}{2} \sum_{j} ||\mathbf{Z}_j||^2_{\mathbf{R}_j} \\
    \nonumber \text{s.t.}\qquad & \mathbf{f}(\boldsymbol{\zeta}^{(j)}_{t+1}, \boldsymbol{\zeta}^{(j)}_{t}, t) = 0 \quad \forall\, t \\
    \nonumber & |\boldsymbol{\zeta}^{(j)}_{t} | \le \boldsymbol{\beta}^{(j)}_t \quad \forall\, t.  
\end{align}
where $||\mathbf{x}||^2_\mathbf{R} := \mathbf{x}^T \mathbf{R} \mathbf{x}$ and $\mathcal{F}(\mathbf{Z}, \mathbf{G}):= |\Tr \{\mathbf{U}_T^\dagger \mathbf{G}\}|$ is the fidelity evaluated by extracting the final unitary, $U_T$, from the optimization variables. The dynamics are denoted implicitly by $\mathbf{f}(\mathbf{Z}_{t+1}, \mathbf{Z}_t, t) = \mathbf{U}_{t+1} - \exp\{-i \Delta t_t \mathbf{H}(\mathbf{a}_t)\} \mathbf{U}_t = 0$; in practice, a series expansion is used to avoid the need to directly compute or differentiate the matrix exponential~\cite{trowbridge2023direct}. The inputs to the unitary smooth pulse problem are the target gate, $\mathbf{G}_j = \mathbf{G}(\boldsymbol{\theta}_j)$, the regularization matrix, $\mathbf{R}_j$, and the bounds, $\mathbf{B}^{(j)} = \begin{bmatrix} \boldsymbol{\beta}^{(j)} _1 & \boldsymbol{\beta}^{(j)} _2 & \cdots & \boldsymbol{\beta}^{(j)} _T \end{bmatrix}$. In practice, we only regularize on the integral of the control and its derivatives, usually as a diagonal matrix across all timesteps, so $||\mathbf{Z}||^2_{\mathbf{R}} = r \sum_t \Delta t_t \left( \mathbf{a}^T_t \mathbf{a}_t + \dot{\mathbf{a}}^T_t \dot{\mathbf{a}}_t + \ddot{\mathbf{a}}^T_t \ddot{\mathbf{a}}_t \right)$ for a scalar regularization parameter, $r > 0$. In addition to regularization, we can set bounds on controls to reflect any constraints set by hardware using $\mathbf{B}$. We can also set the cost to penalize the total time, $\sum_t \Delta t_t$, to realize a minimum time problem~\cite{trowbridge2023direct}. In this way, we can seek out minimum time solutions without relying on a costly second NN, as was suggested by~\cite{sauvage2022optimal}.

\subsubsection{Unitary direct sum problem template} \label{sec:ds}
In order to provide coordinated pulse optimization within Piccolo, we introduce a new unitary direct sum problem template,
\begin{align} \label{eqn:direct-sum-problem}
    \underset{\qquad \mathbf{Z}_{1}, \dots, \mathbf{Z}_N}{\arg \min}&\frac{1}{2} \sum_{\langle i, j \rangle} ||\mathbf{Z}_i - \mathbf{Z}_j||^2_{\mathbf{R}_{ij}} + \frac{1}{2} \sum_{j} ||\mathbf{Z}_j - \mathbf{Z}^{(0)}_j||^2_{\mathbf{R}^{(0)}_j} \\
    \nonumber \text{s.t.}\qquad 
    & |1 - \mathcal{F}(\mathbf{Z}_j, \mathbf{G}_j)| < 10^{-4} \quad \forall\, j \\
    \nonumber & \mathbf{f}(\boldsymbol{\zeta}^{(j)}_{t+1}, \boldsymbol{\zeta}^{(j)}_{t}, t) = 0  \quad \forall\, t,\, j \\
    \nonumber & |\boldsymbol{\zeta}^{(j)}_{t} | \le \boldsymbol{\beta}^{(j)}_t \quad \forall\, t,\, j.
\end{align}
There are a few differences from \eqref{eqn:smooth-pulse-problem}. In particular, fidelity now enters as a constraint (set to $0.9999$), and our cost is now the distance between the optimization variables for nearby problems. We also allow for offset regularizers, $\mathbf{R}^{(0)}_j$, which can be used to keep optimization variables close to initial values, $\mathbf{Z}^{(0)}_j$. These initial values, chosen through Barycentric coordinates, were the key to coordinated re-optimization in~\cite{chadwick2023efficient}. The dynamics of our problem are block diagonal (direct sum), leading to the name of this problem type. It is important to emphasize that we do not have to construct the full direct sum dynamics to store this optimization problem; instead, this problem only scales linearly with the number of sub-problems. It is straightforward to initialize the problem starting from the solutions obtained from \eqref{eqn:smooth-pulse-problem}. 

Permitted by the flexibility of~\eqref{eqn:direct-sum-problem}, we studied many different pairwise regularization variables, including:
\begin{itemize}
    \item[$\circ$] Controls
    \item[$\circ$] Control accelerations
    \item[$\circ$] Unitaries.
\end{itemize}
In Figure~\ref{fig:fig2_compare}, we look more closely at regularization on these three variables. For the purpose of illustration, we use~\eqref{eqn:direct-sum-problem} to synthesize the $\mathbf{R_Z}(\theta)=\exp(-i \theta \mathbf{Z}/2)$ gate using $\mathbf{X}$ and $\mathbf{Y}$ controls (see also, Section~\ref{sec:one-parameter}). We single out a few representative features of the augmented state trajectory in the rows of Figure~\ref{fig:fig2_compare} in order to see how the different regularization schemes change the solution to the optimization problem. The first two rows show the $\mathbf{X}$ drive and corresponding acceleration. The third row shows the trajectory of the observable, $\mathbf{X}$, started from its $-1$ eigenstate and considered in the Heisenberg picture: $\langle \mathbf{X} \rangle = \bra{-} \mathbf{U}(t)^\dagger \mathbf{X} \mathbf{U}(t)\ket{-}$, where $\mathbf{U}(t)$ is the unitary.

We plot these trajectories for a range of reference angles, $\theta$, that have been daisy-chained by regularization using~\eqref{eqn:direct-sum-problem}. All $11$ reference problems are optimized simultaneously (the problem scaling is linear in the size of the subproblem---in this case a single qubit---so this is efficient). To make sure we hit our targets, our objective includes a fidelity constraint of $0.9999$ which is attained by each reference problem for their respective angle. Observe that both control and acceleration-based regularizations achieve similarly tight groupings among reference angles, encouraging good interpolation. Observe that control regularization results in the solution that seems most amenable to linear interpolation, validating the design choice of~\cite{chadwick2023efficient}. We draw attention to the fact that the acceleration-based regularization results in controls that are not as restricted by linear interpolation. We find this to be a benefit because it allows access to control solutions that match intuition, like the use of a zero control for the $\theta=0$ gate. Interestingly, we see that regularizing on unitaries incentivizes the controls to delay any differentiation among trajectories until the last moments. We have seen this feature benefit linear interpolation, but it comes with the price of larger control accelerations in those last moments. 

Ultimately, we choose to align with our NN architecture by regularizing with respect to control accelerations. However, we emphasize that \textit{any} choice of regularization can be combined with our scheme because the control accelerations can always be extracted from the optimization variables of~\eqref{eqn:direct-sum-problem}, no matter how we choose to regularize. We hope that the flexibility offered by~\eqref{eqn:direct-sum-problem} will allow practitioners to make design choices that best align with their circumstances.

\begin{figure}[t]
    \centering
    \includegraphics[width=\columnwidth]{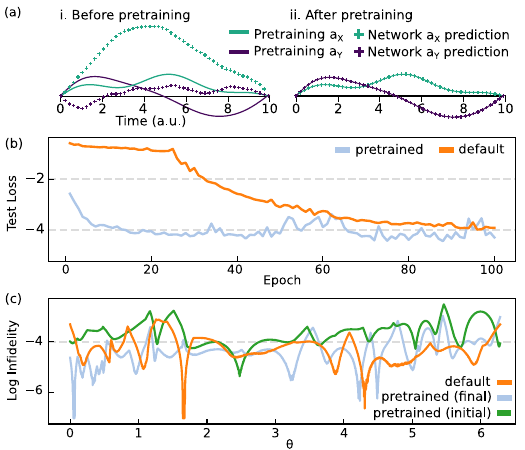}
    \caption{
    Pulses are synthesized for a one-qubit, one-parameter gate family, $\mathbf{R}_\mathbf{Z}(\theta)$, with controls on $\mathbf{X}$ and $\mathbf{Y}$. $\text{(a)}$~A representative pulse from the pretraining data ($\theta=\pi$) is shown. In total, $11$ pretraining angles are chosen (see Figure~\ref{fig:fig2_compare}). $\text{(a) i.}$~Before pretraining, our physics-informed network generates a continuously-differentiable and bounded pulse, and $\text{(a) ii.}$~after pretraining the network has learned to match that pulse to the reference. In $\text{(b)}$, we see that the pretrained network quickly finds an interpolation of the gate family. Test loss reports the average gate fidelity for $4500$ random gate angles. $\text{(c)}$ Log infidelity versus the gate parameter, $\phi$, is shown for the network with and without pretraining. We also include the pretrained (initial), which shows the network after pretraining, but before any fidelity training; the physics-informed network structure leads to good initial interpolation performance.
    }
    \label{fig:fig3_rz_xy}
\end{figure}
\subsection{Pretraining with Piccolo} \label{sec:pretraining}
Reference~\cite{sauvage2022optimal} found that the initial weights of the NN were important for successful training, and overcame this challenge through hyperparameter tuning. Pretraining provides an alternative framework that allows us to start the network close to solutions from optimal control. In this way, the NN spends less resources on discovery of successful controls and focuses more on interpolating the obtained solutions. Indeed, pretraining acts as a form of \textit{active learning}; we provide our learner with an optimal set of reference pulses to bootstrap its training~\cite{cohn1996active}. To select our reference parameters, we use a uniform grid as recommended by~\cite{chadwick2023efficient} 

After selecting reference parameters, we must solve for our pretraining control pulses. We do so by directly coordinating our sampled problems using~\eqref{eqn:direct-sum-problem}. 
Observe that~\eqref{eqn:direct-sum-problem} scales linearly with the number of reference gates we intend to synthesize, so we are limited primarily by the size of the gate's Hilbert space, and not the number of parameters we want to sample. For example, using~\eqref{eqn:direct-sum-problem} in \texttt{Piccolo.jl}~\cite{website2024piccolo}, we were able to set up and solve a direct sum problem involving $100$ qubits on a laptop in $<10$ minutes. One advantage of combining our reference parameters into a single optimization problem is that we can directly probe the minimum time needed to achieve the gate family. We do so by taking the mutual problem timesteps, $\boldsymbol{\Delta}\mathbf{t} = \begin{bmatrix} \Delta t_1 & \Delta t_2 & \cdots & \Delta t_T \end{bmatrix}$, as part of the augmented controls, through which we can introduce a minimum time cost to the problem by penalizing $\sum_t \Delta t_t$~\cite{trowbridge2023direct}. As problem sizes grow, a direct sum problem for an ensemble of gates becomes memory limited, and heuristics become key. 

In our experiments, pretraining data is generated by using regularization on acceleration values (see Figure~\ref{fig:fig2_compare}). Regularization forces nearby gate parameters to have similar accelerations, which is desirable for pretraining because the NN output (accelerations) is a continuous function of the network input (parameters). Recall (Section~\ref{sec:ds} and Figure~\ref{fig:fig2_compare}) that acceleration regularization is not as amenable to linear interpolation as control regularization; however, with a NN as interpolant, this is not an issue. The loss function for pretraining is simply the mean squared error between the output of the network at a sampled parameter value and the corresponding pretraining pulse. We show the effect of pretraining in Figure~\ref{fig:fig3_rz_xy}~(a).

\subsection{Interpolation using neural networks}  \label{sec:training}
To teach our network to interpolate over the gate parameters, we choose our network loss to be the sum of the gate infidelities in a training batch:
\begin{equation} \label{eqn:nn_loss}
    \mathcal{L}(\mathbf{W}){=} \sum_{j\in\text{Batch}} 1 - \mathcal{F}\left(\mathbf{U}_T(\ddot{\mathbf{a}}(\boldsymbol{\theta}_j; \mathbf{W}))^\dagger \mathbf{G}(\boldsymbol{\theta}_j) \right)
    + \lambda \norm{\ddot{\mathbf{a}}(\boldsymbol{\theta}_j; \mathbf{W})}_1
\end{equation}
where $\mathbf{W}$ are the network weights, $\mathbf{U}_T$ are the dynamics at the final time, and $\mathbf{G}$ is the parameterized gate. To regularize our solutions and enforce sparsity on the accelerations output by the network, we add an $L1$ norm. We train our network over multiple epochs. In each epoch, we select $500$ gate parameters, $\boldsymbol{\theta}_j$. We train in batches of $50$. After exhausting the samples contained in the current epoch, we select a new set of $500$ samples for the next epoch. In this way, we teach the network to extrapolate to the entire gate family.

Notice that~\eqref{eqn:nn_loss} is actually an indirect approach to optimal control because the dynamics are enforced at all timesteps. Usually, indirect approaches to quantum optimal control are solved using piecewise constant controls or by expanding controls in a linear function basis, which allow for efficient numerical gradients~\cite{defouquieres2011second}. Here, we rely on automatic differentiation to optimize~\eqref{eqn:nn_loss}, with the control represented by our physics-informed NN.

The reference parameters we use to train our NN are decided on the fly. In an attempt to exploit this freedom, we again looked to active learning. This time, we sought to bias our samples at each epoch based on the current network status~\cite{lewis1995sequential}: we applied standard importance sampling with respect to infidelity to realize the next set of examples at each epoch, hoping to preference locations in gate parameter space that were underperforming. Ultimately, we found this approach offered a small and inconsistent advantage relative to random training sets on our examples, so we select our epoch samples randomly.

\subsection{Calibration via transfer learning} \label{sec:calibrating}
The gates or pulses that a user designs using models will not directly agree with their implementation on hardware. Assuming there is always some model imperfections or uncertainty, \textit{calibration} is necessary to force the hardware results to fall into line with the designed operations. Common causes for recalibration include qubit frequency drift due to environmental changes, spurious couplings to other systems, and distortions or crosstalk across control lines; in summary, models of a quantum computer are always fluctuating over time in small ways, so that consistent characterization and recalibration are a daily fact of life. Various approaches exist for calibration, which depend on the needs of the application, but can include error amplification~\cite{sheldon2016characterizing}, closed-loop optimization~\cite{egger2014adaptive,kelly2014optimal}, and contextual, system-wide optimization~\cite{arute2019quantum,klimov2020snake}. 

Directly optimizing our NN cost using closed-loop fidelity measurements (similar to~\cite{egger2014adaptive,kelly2014optimal}) would be prohibitive. However, our framework allows us to move between the training and pretraining phases of our algorithm with the same network. By applying our pretraining stage directly to calibrated pulses, we can efficiently apply \textit{transfer learning} to train our network to adapt to the new context. In prior work, reinforcement learning agents learned to improve the fidelity of a two-qubit cross-resonance gate, and it was shown that the agent could be efficiently recalibrated by training the model with new experimental data~\cite{baum2021experimental}. Similarly, our network retains information about the structure of the gate family it has interpolated, so that a few representative examples allow us to recalibrate the entire surface. Moreover, our scheme accommodates whatever calibration routine is efficient for the device, as long as the routine can be applied to realize calibrations of a few reference pulses of the calibrated gate family.

\begin{figure*}[t]
    \centering
    \includegraphics[width=\textwidth]{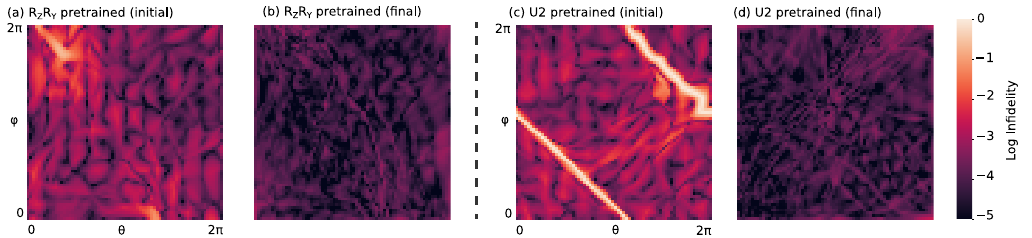}
    \caption{Heatmaps plotting the log of the infidelity for two parameter gate families $\mathbf{R}_\text{Z}(\theta) \mathbf{R}_\text{Y}(\phi)$~\eqref{eqn:rzry} and \textbf{$\mathbf{U}2((\theta, \phi)$}~\eqref{eqn:u2}. The gate $\mathbf{R}_\text{Z} \mathbf{R}_\text{Y}$ is shown in the first two panels, (a) after pretraining but before training, and (b) after training. Pretraining provides an excellent initial guess (see Table~\ref{tab:table1}). The last two panels show the same results for \textbf{$\mathbf{U}2$}; in this case, lines of high infidelity appear at $\theta + \phi = \pi, 3\pi$, cutting the torus into two regions and hurting the advantage pretraining offers to global pretraining in Table~\ref{tab:table1}. In either case, both gate families converge to satisfactory fidelity values after training.}
    \label{fig:fig4_u2}
\end{figure*}
\section{Pulse-level VQE gate families} \label{sec:synth}
In this section, we synthesize various examples of continuous families of quantum gates having one and two parameters. To explore pulse-level synthesis, we use a model~\eqref{eqn:model} motivated by superconducting transmon systems in the qubit approximation. In transmons, a pairwise interaction can be engineered which leads to the swapping of excitations between two qubits, $\mathbf{H}^\text{iSWAP}_{12} = \frac{1}{2}(\mathbf{X}^{(1)}\mathbf{X}^{(2)} + \mathbf{Y}^{(1)} \mathbf{Y}^{(2)}$). Here, $\mathbf{X}^{(j)}$ denotes an operator acting on qubit $j$. The rotation about $\mathbf{H}_\text{iSWAP}$ leads to the native two-qubit gate known as iSWAP, hence the naming. The iSWAP interaction can be parametrically achieved by bringing capacitively coupled flux-tunable qubits into resonance, or by directly modulating a tunable coupling element, among other approaches~\cite{krantz2019quantum}. Upon also considering microwave control lines driving individual qubit rotations, we have
\begin{equation} \label{eqn:model}
    \mathbf{H}(\mathbf{a}_t) = \sum\limits_j \text{a}_{\text{X}j,t} \mathbf{X}^{(j)} + \text{a}_{\text{Y}j,t} \mathbf{Y}^{(j)} 
    + \sum\limits_{\langle j, k \rangle} \text{a}_{jk,t} \mathbf{H}^\text{iSWAP}_{jk}
\end{equation}

Our choice of model is consistent with experimental work on parameterized gate families~\cite{abrams2020implementation,foxen2020demonstrating} and captures the effective interaction that emerges when considering the models in~\cite{sauvage2022optimal,chadwick2023efficient}.  We choose not to directly model access to the $\mathbf{Z}$ control (flux tunability) for each qubit; this is to make sure control is hard enough. As control gets easier, so does interpolation (Appendix~\ref{sec:qc:theory}).

For each of the continuously-parameterized gate families we study, we report the average gate fidelity and standard deviation of gate fidelity over a densely sampled test set of gate parameters. The mean and standard deviation of average gate fidelity for these test sets are summarized in Table~\ref{tab:table1}, alongside training efficiency. To quantify the training efficiency, we report the number of epochs to exceed a threshold average gate fidelity,~$0.9999$. Recall that an epoch corresponds to $500$ training samples, so we can multiply this by the number of epochs to determine the number of unique gate parameters drawn to achieve interpolation. 
The key takeaway for this table is that pretraining leads to fidelities that are as good as default NN interpolation for all gates we consider, while reducing the number of epochs needed to reach the threshold fidelity. More information about each gate family can be found in the referenced section. 

\begin{table}[t]
    \centering
    \begin{tabular}{p{1.25cm}|p{1.25cm}|p{1.0cm}|p{1.0cm}|p{1.0cm}}
        \multicolumn{1}{c}{} & \multicolumn{1}{l}{} & \multicolumn{1}{l}{} & \multicolumn{1}{l}{} & \multicolumn{1}{l}{Epochs to} \\
        \multicolumn{1}{c}{Gate} & \multicolumn{1}{l}{Network} & \multicolumn{1}{l}{Mean} & \multicolumn{1}{l}{Std. dev.} & \multicolumn{1}{l}{reach threshold} \\
    
        \hline
        \multirow{2}{*}{$\mathbf{R}_\text{Z}$~\eqref{eqn:rz}} & Pretrained   & 8.10e-05 & 1.21e-4 &  14\\
        \cline{2-5}
        & Default & 9.16e-5 & 1.29-4 & 100\\
        \hline
    
        \multicolumn{5}{l}{} \\ 
        \hline
        \multirow{2}{*}{$\mathbf{R}_\text{Z} \mathbf{R}_\text{Y}$~\eqref{eqn:rzry}} & Pretrained   & 9.69e-5 & 1.07e-4 & 437 \\
        \cline{2-5}
        & Default   & 9.33e-5 & 1.13e-4 & 731 \\
        \hline
        
        \multicolumn{5}{l}{} \\ 
        \hline
        \multirow{2}{*}{\textbf{$\mathbf{U}2$}~\eqref{eqn:u2}} & Pretrained & 9.77e-5 & 1.22e-4 & 2246$^*$\\   
        \cline{2-5}
        & Default  & 8.62e-5 & 1.28e-4 & 178\\ 
        \hline
    
        \multicolumn{5}{l}{} \\ 
        \hline
        \multirow{2}{*}{$\widetilde{\mathbf{A}}_{12}$~\eqref{eqn:adapt2}} & Pretrained   & 9.05e-5 & 1.20e-4 & 682 \\
        \cline{2-5}
        & Default   & 9.92e-5 & 9.72e-5 & 662 \\
        \hline
        \multicolumn{5}{l}{} \\ 
        \multicolumn{4}{l}{} & \multicolumn{1}{l}{\footnotesize{$^*$see~\ref{sec:two-parameter}}}  \\

    \end{tabular}
    \caption{Average gate infidelity \& training efficiency}
    \label{tab:table1}
\end{table}

All pretraining data was generated on a laptop with an Intel Core i7 processor ($2.20$~GHz, $12$~cores) and $16$~GB RAM. Training was performed using basic-plan GPU and CPU resources from Google Colab.

\subsection{One-parameter gates}  \label{sec:one-parameter}
The simplest non-trivial example of a continuously-parameterized gate family is a one-parameter, one-qubit gate based on a rotation that cannot be directly implemented using the available control Hamiltonians. For this purpose, we implement a phase gate,
\begin{equation} \label{eqn:rz}
    \mathbf{R}_\text{Z}(\theta) = \begin{pmatrix} e^{-i \frac{\theta}{2}} & 0 \\ 0 & e^{i \frac{\theta}{2}} \end{pmatrix},
\end{equation}
which requires coordinating the $\mathbf{X}$ and $\mathbf{Y}$ control available in our model.

In Figure~\ref{fig:fig3_rz_xy}, we plot the result of applying our framework to~\eqref{eqn:rz}. We select $11$ pretraining points and solve~\eqref{eqn:direct-sum-problem} to coordinate their optimal control solutions. We find that pretraining offers a $7\times$ increase in training efficiency, as measured by the time it takes the NN to exceed a threshold average gate fidelity of~$0.9999$. The default NN uses random weights; we select a representative example, which includes a characteristic early plateau during which the network learns good control solutions before finding interpolations. For this simple example, observe that the NN after pretraining is already at $99.9$ fidelity for most parameters. Combining this NN with training using~\ref{eqn:nn_loss} is sufficient to smooth over the remaining parameters in just 14 training iterations (Table~\ref{tab:table1}). We expect trade-offs between the optimal control complexity due to pretraining samples and the required number of network training samples, but a full sensitivity analysis is beyond the scope of this work.

\subsection{Two-parameter gates} \label{sec:two-parameter}
In Figure~\ref{fig:fig4_u2}, we show the results of applying our framework to interpolate two-parameter gate families. We consider a pair of single qubit families,
\begin{equation} \label{eqn:rzry}
    \mathbf{R}_\text{Z}(\theta)\mathbf{R}_\text{Y}(\phi) = \exp(-i \frac{\theta}{2} \mathbf{Z}) \exp(-i \frac{\phi}{2} \mathbf{Y})
\end{equation}
and
\begin{equation} \label{eqn:u2}
    \mathbf{U}2(\theta, \phi) = \frac{1}{\sqrt{2}} \begin{pmatrix} 1 & -e^{i\phi} \\  e^{i\theta} & e^{i(\theta + \phi)} \end{pmatrix}.
\end{equation}

In Figure~\ref{fig:fig4_u2}, we show the results of our framework for our two-parameter gate families. We plot the infidelity as a heatmap over the parameters in $[0,2\pi] \cross [0,2\pi]$; in particular, we single out the initial network infidelity after completing pretraining, and the final infidelity after completing both pretraining and training. Not shown is the random network after training because it is comparable to the pretrained network after training.

We generate pretraining data by solving~\eqref{eqn:direct-sum-problem} with $11\times11=121$ equally-spaced lattice points. Comparing Figure~\ref{fig:fig4_u2}(a) with Figure~\ref{fig:fig4_u2}(c), observe that pretraining behaves differently for the two gate families. For~\eqref{eqn:rzry}, pretraining performance behaves in an approximately uniform way. In Table~\ref{tab:table1}, we report a clear $2\times$ efficiency for the pretrained network. Contrast this with~\eqref{eqn:u2}, where high-fidelity lines appear at $\theta + \phi = \pi$ and $3\pi$---dividing the torus into two regions. These barriers are due to the minimum time constraint we impose on the problem; the two regimes emerge due to optimal controls taking different incompatible shortest paths on the Bloch sphere. During optimization, each reference parameter converges to one of the two regions, with the divisions at $\pi$ and $3\pi$ the furthest rotations from identity. A correlated feature of this division in reference parameters is a loss in training efficiency for the pretrained network, which must find a way to remap incompatible controls to interpolatable values. Observe in Table~\ref{tab:table1} that the pretrained network has a significantly worse training efficiency when compared to the random network. The pretrained NN's lack of training advantage for a global interpolation solution is not surprising given Figure~\ref{fig:fig4_u2}, and an advantage could be restored by using two different interpolators, one for each of the two regions.  We emphasize that the NN is still able to converge to a good gate family even if pretraining is not optimal---one of the principal reasons we choose to use a NN over linear interpolation is the universal approximation guarantees are helpful for hard control problems where the control landscape can be nonlinear.

\subsection{Advantages of gate families in VQE circuits} \label{sec:advantages}

\begin{figure}[t]
    \centering
    \includegraphics[width=\columnwidth]{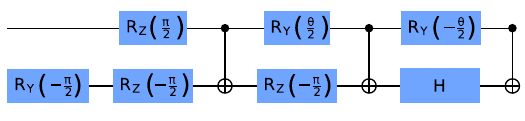}
    \caption{Two-qubit QEB-ADAPT-VQE operator, $\widetilde{\mathbf{A}}_{12}(\theta)$~\eqref{eqn:adapt2}~\cite{yordanov2020efficient,Yordanov_2021}.
    }
    \label{fig:circuit}
\end{figure}
\subsubsection{QEB-ADAPT-VQE (Two qubits, one parameter)}
We also demonstrate the performance of our approach using a two-qubit gate. When using \textsc{QEB-ADAPT-VQE}, the two qubit gate 
\begin{equation} \label{eqn:adapt2}
    \widetilde{\mathbf{A}}_{12}(\theta) = \exp(i\frac{\theta}{2}\left(\mathbf{X}_{1}\mathbf{Y}_{2}-\mathbf{Y}_{1}\mathbf{X}_{2}\right) )
\end{equation}
occurs in the ansatz built. When decomposed into a basis gate set, this gate contains $3$ CNOT gates (Figure~\ref{fig:circuit}). 

We can derive a baseline CNOT gate for our model~\eqref{eqn:model} using the \textit{mintime problem template} in \texttt{Piccolo.jl}~\cite{website2024piccolo,trowbridge2023direct}. A mintime problem can be obtained from~\eqref{eqn:smooth-pulse-problem} by adding timesteps as an additional control variable in the augmented control. Then, the cost is replaced with a time cost, $\sum_t \Delta t_t$, and the fidelity is moved to a constraint. This new problem is now enabled to directly seek mintime up to some desired fidelity tolerance. To compare our pulse-level solution to the standard gate set, we synthesize a CNOT gate up to a $0.9999$ fidelity tolerance according to the problem constraints of our model (which includes amplitude and acceleration bounds set to $1\si{GHz}$ and $1\si{GHz^3}$, respectively). For these constraints, the CNOT duration we obtained is $\sim 5.95\si{ns}$. If we use virtual $\mathbf{Z}$ rotations, we can approximate the total pulse schedule time to be these three CNOTs plus two $\pi/2$-Pauli rotations (when $\theta=0$) and $3$ $\pi/2$-Pauli rotations (when $\theta=\pi$). Solving a minimum time problem on a $\mathbf{R}_\text{X}(\pi/2)$ gate for our problem constraints results in a duration of $\sim5\si{ns}$. The inclusion of these rotation gates results in a pulse schedule that ranges from $27.8\si{ns}$ to $32.8\si{ns}$, depending on the value of $\theta$. On the other hand, when we used our NN approach to directly compile the $\widetilde{A}_{12}$ gate, we achieved a comparable fidelity using only $~10\si{ns}$. In the case of virtual gates this is an improvement of $2.78-3.28\text{x}$ depending on $\theta$. Because of the tradeoff between time and frequency units, these results can be rescaled to reflect appropriate amplitude bounds for any hardware platform, with the improvement ratio a unitless quantity that should be preserved by this conversion.

Due to the increased size of the Hilbert space, our pretraining scheme failed to provide a significant advantage on this problem (Table~\ref{tab:table1}). We expect a more sophisticated pretraining protocol will likely be able to recover the advantage seen in the single qubit gates (see Section~\ref{sec:future}).

\begin{figure}[t]
    \centering
    \includegraphics[width=\columnwidth]{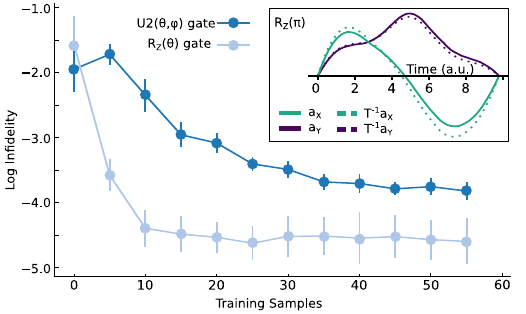}
    \caption{
    The average infidelity of a densely sampled test set is plotted versus the number of calibrated training samples used during transfer learning for gate families~\eqref{eqn:rz} and~\eqref{eqn:u2}. Initial infidelities are high because of the modeled distortions from device drift, which transfer learning overcomes by following the protocol of pretraining, but now using calibrated controls and starting from a previously trained network. Error bars show uncertainty over an ensemble of pulse-distorting transfer matrices, $\mathbf{T}$. ($\text{inset}$) Shown in solid are the two control pulses from the original network, which implement $\mathbf{R}_\mathbf{Z}(\pi)$. The dashed pulses are the calibrated controls, and they correct for the transfer $\mathbf{a}_t = \mathbf{T}(\mathbf{T}^{-1} \mathbf{a}_t)$. These calibrated controls are fed to the network as the training samples.
    }
    \label{fig:fig5_cal}
\end{figure}
\subsection{Calibration for control drift}
We consider a model of pulse distortion between the AWG and the qubits for the purpose of studying transfer learning of our NN. Our simplified model includes asymmetric classical crosstalk between different control lines and amplitude uncertainty on axis~\cite{chen2018metrology}. In sum, we consider control \textit{transfer matrices} $\mathbf{a}'_t = \mathbf{T}\mathbf{a}_t$, where
\begin{equation}
    \mathbf{T} = \begin{pmatrix} 1 + n_{11} & n_{12} \\ n_{21} & 1 + n_{22} \end{pmatrix}
\end{equation}
for $n_{ij} \sim \mathcal{N}(0, 0.10)$. These values reasonably capture the noise models described in~\cite{chen2018metrology} and capture the upper bound of the day-over-day changes in $\mathbf{X}$-gate drive amplitudes, as accessed from the calibration histories of $127$-qubit fixed-frequency, fixed coupler systems deployed by IBM Quantum~\cite{website2024ibm}.

In order to study the ability of our framework to adapt to this new regime, we apply transfer learning: First, the parameters of a NN predicting controls for a gate family are frozen, except for the last layer. Next, we sequentially choose reference parameters $\boldsymbol{\theta}_j$. For each parameter, we obtain a calibrated control matrix, $\mathbf{T}^{-1}\mathbf{a}(\boldsymbol{\theta}_j)$\footnote{
    The calibrated control matrix can be obtained by an experimentalist using whatever tools is most practical. We expect our physics-informed network with continuously differentiable controls to provide easy-to-calibrate pulse shapes, relative to piecewise-constant controls.
}.
Our approach is to train the network to predict each new training point until the gradients saturate, and then continue with the next point. Explicitly, the stopping condition used is the norm-squared of the gradient---it is computed after every epoch, and if it ever falls below a certain threshold, the model stops training.

In Figure~\ref{fig:fig5_cal}, we show the result of applying our transfer learning scheme for a one and two-parameter gate family, $\mathbf{R}_\mathbf{Z}(\phi)$ and $\mathbf{U}2(\phi, \lambda)$, respectively. In the figure, we see that the transfer matrix does meaningfully distort the controls, and the average gate family fidelity has suffered---the new fidelity range is $0.9-0.99$ instead of the sub-$0.9999$ that we found upon completion of our original training framework. We observe that for the one-parameter gate, $<10$ controls are needed to reach $0.9999$ fidelity---fewer than the original presample of $11$, and much less than the data required by the training stage. The two-parameter gate also uses far fewer parameters than pretraining, converging to an average gate fidelity of $0.999$ in $<25$ iterations.

Because we freeze all but the last layer of our network, the only changes we apply are to the linear weights and biases of the final linear network layer. Observing that our transfer matrix is a linear transformation of the controls, we note that we should always be able to retain the original infidelity by applying the inverse of this transformation to the network's last layer of parameters. As linear transfer matrices are a common way to model control distortion, we expect this approach to be efficient in many practical settings. If a nonlinear \textit{transfer function} is applied to the controls, then we can easily add activation to our final transfer layer, likely at the price of training efficiency.

\section{Conclusion} \label{sec:future}
We have introduced a framework for continuously-parameterized families of quantum gates, with a particular focus on VQE applications. We showed for \textsc{QEB-ADAPT-VQE} that continuously-parameterized gate families could be used to achieve a speedup of up to $\mathbf{3.28\times}$. We offer new explanations for the success of linear and nonlinear interpolation of continuous-gate families, based on bilinear control theory. Our framework builds on the advantages of past work, and overcomes any limitations by taking a joint approach. By integrating quantum optimal control with physics-informed machine learning, we generate control surfaces that seamlessly interpolate among these gate families. This allows for expressiveness beyond linear interpolation, which is crucial for hard quantum control tasks like mintime or limited control. Our efficient synthesis approach---guided by optimal control for pretraining and active learning---allows us to rapidly converge to nonlinear control surfaces. We observe that the more constrained the optimization, the more important the initial optimal control guidance becomes important, so that we expect substantial benefits by establishing better integration of coordinated interpolation for problems that must be broken up across multiple optimizations. Through simulations, we have demonstrated the effectiveness of our approach by synthesizing relevant one and two
qubit gates with one or two parameters. We find speedups in our study. Overall, our framework showcases the practical integration of optimal control tools with machine learning, providing valuable enhancements to current quantum algorithms. By introducing the possibility of efficient recalibration of gate families for a device noise modeled by control transfer matrices, we expect our work to help enable practical utilization of gate families for advantage beyond standard gate sets.

Regarding future work, we suggest the following items:
\begin{itemize}
    \item[$\circ$] \textit{Coordinated optimization heuristics:} We expect that more efficient pretraining strategies are essential for maximizing the advantage to training efficiency as the system size increases. With larger initial problem sizes, global coordination of reference parameters using~\eqref{eqn:direct-sum-problem} becomes infeasible, and good coordinated optimization heuristics like~\cite{chadwick2023efficient} are needed which efficiently capture the weak long range correlations among distant reference parameters. Alternatively, regions like those seen for $U2$ in Figure~\ref{fig:fig4_u2} should be identified and exploited to push the advantage of pretraining for all hard quantum control problems. Advantageously, as problem size and control complexity increase, so do the number of local minima, and we expect the possible advantage from pretraining to increase---justifying the extra effort.
    \item[$\circ$] \textit{Alternative network architectures:} In this work, we used a simple feedforward NN to map between piecewise constant accelerations. We expect that NNs appropriate for discrete time series, such as the RNN family of architectures, could be adapted to help reduce the prediction space by making a better account of time correlations via the network structure. Reinforcement learning (e.g.~\cite{baum2021experimental}) can also be combined with our pretraining scheme; gate parameters can be stored as environment variables, and policy learning for an agent might be better adapted to larger action spaces than our naive NN.
    \item[$\circ$] \textit{Hardware demonstrations:} Efficient gates often require modeling interactions with more than two levels. Including higher level gates and subspace targets to address leakage or realize $CZ$ gates will allow for more realistic gate synthesis, moving our experiments closer to hardware realizations. With good system knowledge, the control design can provide a strong starting point, from which transfer learning can be used to test NN calibration.
    \item[$\circ$] \textit{Families of calibration parameters:} Reference~\cite{sauvage2022optimal} suggested adding gate parameters based on intrinsic system parameters or external environmental factors. Such gate families are defined over the space of uncertain parameters, and this type of family could be integrated with transfer learning for calibration.
\end{itemize}

\subsection{Code availability}
Our framework, in particular~\eqref{eqn:direct-sum-problem}, is implemented using a fully open-source control backend~\cite{website2024piccolo}, so that users can contribute and adapt the work to their needs. We hope that our approach enables many of the innovations suggested in the future work in~\cite{chadwick2023efficient}. We expect benefits to be had by exploring other graph-based optimal control problems, which lend naturally to quantum computing applications with local device connectivity. The linear scaling of~\eqref{eqn:direct-sum-problem} means that we are limited primarily by the size of the Hilbert space, not the number of problems in the regularization graph (i.e. not by the reference gate parameters)---we found that problems involving $100$ qubits could be optimized in minutes on a laptop.

We additionally share a GitHub repository with an example of our pretraining, training, and calibration frameworks in Google Colab notebooks at~\url{https://github.com/BBhattacharyya1729/vqe-gate-families}.

\section*{Acknowledgments} \label{sec:ackn}
The authors would like to thank Jason Chadwick for discussions about the implementation of~\cite{chadwick2023efficient}. 
This research was supported by EPiQC, an NSF Expedition in Computing, under award CCF-1730449; STAQ under award NSF Phy-1818914; the US Department of Energy Office of Advanced Scientific Computing Research, Accelerated Research for Quantum Computing Program; the NSF Quantum Leap Challenge Institute for Hybrid Quantum Architectures and Networks (NSF Award 2016136); and the U.S. Department of Energy, Office of Science, National Quantum Information Science Research Centers. FTC is Chief Scientist for Quantum Software at Infleqtion and an advisor to Quantum Circuits. AJG was supported by an appointment to the Intelligence Community Postdoctoral Research Fellowship Program at University of Chicago administered by Oak Ridge Institute for Science and Education (ORISE) through an interagency agreement between the U.S. Department of Energy and the Office of the Director of National Intelligence (ODNI).

\appendices
\renewcommand{\theequation}{A\arabic{equation}}
\numberwithin{equation}{section}

\section{Linear and nonlinear interpolation} \label{sec:qc:theory}
Rotation gates are the most familiar example of interpolatable gate families. Consider a one parameter gate family generated by rotation, $\mathbf{G}(\theta) = \exp(-i \theta \mathbf{H}_\text{eff.})$, and suppose that our effective Hamiltonian can be expressed using the provided control Hamiltonians, $\mathbf{H}_\text{eff.} = \sum_j c_j \mathbf{H}_j$. The simplest case for analysis is constant control. Consider $\text{a}_j(\theta) = \theta c_j/T \Delta t$ for gate time $T \Delta t$. The dynamics are
\begin{equation} \label{eqn:linear}
    \mathbf{U}(T) = e^{-i \mathbf{H}(\mathbf{a}(\theta)) T \Delta t} = e^{-i \mathbf{H}(\frac{\theta}{T \Delta t}\mathbf{c}) T \Delta t} = e^{-i \theta \mathbf{H}_\text{eff.}},
\end{equation}
so that the entire family of gates can be found by \textit{linearly} rescaling the controls,
$\mathbf{H}(\mathbf{a}(\theta)) \approx \theta \mathbf{H}(\mathbf{c}/T \Delta t)$.

Now, for the same rotation, suppose that we aren't as lucky and require the first nontrivial element of the Lie algebra~\cite{dalessandro2021introduction}: $\mathbf{H}_\text{eff.} = a [\mathbf{H}_1, \mathbf{H}_2]$. For example, this includes the case of synthesizing an $\mathbf{R_Z}(\theta)=\exp(-i \theta \mathbf{Z} / 2)$ gate using $\mathbf{X}$ and $\mathbf{Y}$ control. If controls are sufficiently small, we can take the \textit{Magnus expansion} to express the time independent generator of the rotation using the control Hamiltonians, $\mathbf{H}_\text{eff.} \overset{!}{=} \sum_n \mathbf{H}^{(n)} \approx \mathbf{H}^{(1)} + \mathbf{H}^{(2)}$ where 
\begin{align}
    \mathbf{H}^{(1)} &= \frac{1}{T} \sum\nolimits_{t=1}^T \mathbf{H}(\mathbf{a}_t) \\  \nonumber
    \mathbf{H}^{(2)} &= \frac{-i \Delta t}{2 T} \sum\nolimits_{t=1}^T \sum\nolimits_{t'=1}^t [\mathbf{H}(\mathbf{a}_t), \mathbf{H}(\mathbf{a}_{t'})].
\end{align}
For our example of $\mathbf{R_Z}(\theta)$, these equations represent necessary conditions on the controls. Setting $\mathbf{H}(\mathbf{a}_t(\theta)) = a_{\text{X}t}(\theta)\mathbf{X} + a_{\text{Y}t}(\theta)\mathbf{Y}$ and $\mathbf{H}_\text{eff.} = \mathbf{Z}$, we find
\begin{align} \label{eqn:nonlinear}
    0 \overset{!}{=} &\sum\nolimits_{t=1}^T a_{\text{X}t} = \sum\nolimits_{t=1}^T a_{\text{Y}t} \\  \nonumber
    \frac{\theta}{2} \overset{!}{=}  &\frac{\Delta t}{T} \sum\nolimits_{t=1}^T \sum\nolimits_{t'=1}^t a_{\text{X}t} a_{\text{Y}t'} - a_{\text{X}t'} a_{\text{Y}t} .
\end{align}
Unlike~\eqref{eqn:linear}, there is now a \textit{nonlinear} relationship between the controls and the gate angle, $\mathbf{H}(\mathbf{a}_t(\theta)) \ne \theta \mathbf{H}_t$.

In practice, we can treat the full unitary dynamics piecewise, such that $\mathbf{U}(T, 1) = \prod_{t=1}^{T-1} \mathbf{U}(t{+}1, t)$ is a time-ordered product, and $\mathbf{U}(t{+}1, t) = \exp(-i \mathbf{H}(\mathbf{a}_t(\theta))\Delta t)$. Considering each timestep in the manner of~\eqref{eqn:linear}, we see that linear interpolation can be approximately achieved if the control dependence remains approximately linear, $\mathbf{H}(\mathbf{a}_t(\theta)) \approx \theta \mathbf{H}_t$. When possible, this feature is attained numerically by solving optimal control problems like~\eqref{eqn:direct-sum-problem}.

\newpage
\bibliographystyle{IEEEtran}
\bibliography{universal}

\end{document}